\documentclass[11pt,twoside]{article}
\usepackage{asp2014}
\aspSuppressVolSlug
\resetcounters
\begin{document}

\title{Quantifying \& Mitigating Satellite Constellation Interference with SatHub}

\author{Meredith L. Rawls,$^{1,2}$ Constance E. Walker,$^{3,2}$ Michelle Dadighat,$^{3,2}$ Harrison Krantz,$^{4,5,2}$ Siegfried Eggl,$^{6,2}$ and Mike Peel$^{7,2}$}
\affil{$^1$University of Washington, Seattle, WA, USA; \email{mrawls@uw.edu}}
\affil{$^2$IAU Centre for the Protection of the Dark and Quiet Sky from Satellite Constellation Interference (CPS)}
\affil{$^3$NSF's NOIRLab, Tucson, AZ, USA}
\affil{$^4$Applied Research Associates Inc., Albuquerque, NM, USA}
\affil{$^5$University of Arizona Steward Observatory, Tucson, AZ, USA}
\affil{$^6$University of Illinois at Urbana-Champaign, Champaign, IL, USA}
\affil{$^7$Imperial College London, Blackett Lab, London, UK}

\paperauthor{Meredith L. Rawls}{mrawls@uw.edu}{0000-0003-1305-7308}{University of Washington}{Department of Astronomy/DiRAC/Vera C. Rubin Observatory}{Seattle}{WA}{98195}{USA}
\paperauthor{Constance E. Walker}{connie.walker@noirlab.edu}{0000-0003-0064-4298}{NSF's NOIRLab}{}{Tucson}{AZ}{85719}{USA}
\paperauthor{Michelle Dadighat}{michelle.dadighat@noirlab.edu}{}{NSF's NOIRLab}{}{Tucson}{AZ}{85719}{USA}
\paperauthor{Harrison Krantz}{harryk@arizona.edu}{0000-0003-0000-0126}{University of Arizona Seward Observatory}{}{Tucson}{AZ}{85719}{USA}
\paperauthor{Siegfried Eggl}{eggl@illinois.edu}{0000-0002-1398-6302}{University of Illinois at Urbana-Champaign}{Department of Aerospace Engineering/Department of Astronomy}{Champaign}{IL}{61820}{USA}
\paperauthor{Mike Peel}{m.peel@imperial.ac.uk}{0000-0003-3412-2586}{Imperial College London}{Blackett Lab}{London}{}{SW7 2AZ}{UK}


\begin{abstract}
This Birds-of-a-Feather (BOF) session on 6 November 2023 was organized by leaders and members of SatHub at the International Astronomical Union Centre for the Protection of the Dark and Quiet Sky from Satellite Constellation Interference (IAU CPS). SatHub is dedicated to observations, data analysis, software, and related activities. The session opened with a talk on the current state of affairs with regards to satellite constellation mitigation, with a focus on optical astronomy, and moved to focused discussion around the top-voted topics. These included tools and techniques for forecasting satellite positions and brightnesses as well as streak detection and masking.
\end{abstract}


\section{Introduction}
SatHub is a community-driven hub that supports open collaboration on astronomy-related studies of commercial satellite constellations, their impact on our sky, the efficacy of various mitigation schemes, and related topics. It is part of the IAU CPS and was first conceived by \citet{rawls21-satcon2}.

A significant focus of SatHub is to disseminate and support development of software tools to aid in the quantification and mitigation of satellite constellations, closely following both \citet{rawls21-satcon2} and \citet{mcdowell21-satcon2}. The session therefore opened with a summary of the current state of affairs. Figure \ref{fig1} shows four different ways to characterize the present situation as a function of time. Each plot illustrates how the number of objects in orbit is increasing significantly faster than ever before. With respect to astronomy and the interests of SatHub is the proliferation of bright, commercial low-Earth orbit satellites that create a new source of globally visible light pollution.

In order to quantify the impact of this rapidly changing low-Earth orbit environment on astronomical science, it is necessary to get a handle on several imperfectly-known quantities: the number of satellites and their orbits, the satellite brightness distribution, various instrument response functions, and the effects of data processing pipelines. Much initial work has been done in each of these areas \citep[see, e.g.,][]{krantz23,fankhauser23,hu22,tyson20,hasan22,lawler22}. However, the fact remains that the critical quantities are often time-variable and at best challenging to predict.

Therefore, following successful SATCON1, SATCON2 \citep{walker20-satcon1,hall21-satcon2}, and Dark \& Quiet Skies I \& II \citep{walker20-dqs1,walker21-dqs2} workshops in 2020 and 2021, the IAU CPS and its four hubs were established in 2022 to bring together experts to address this problem from multiple angles.

\articlefigure{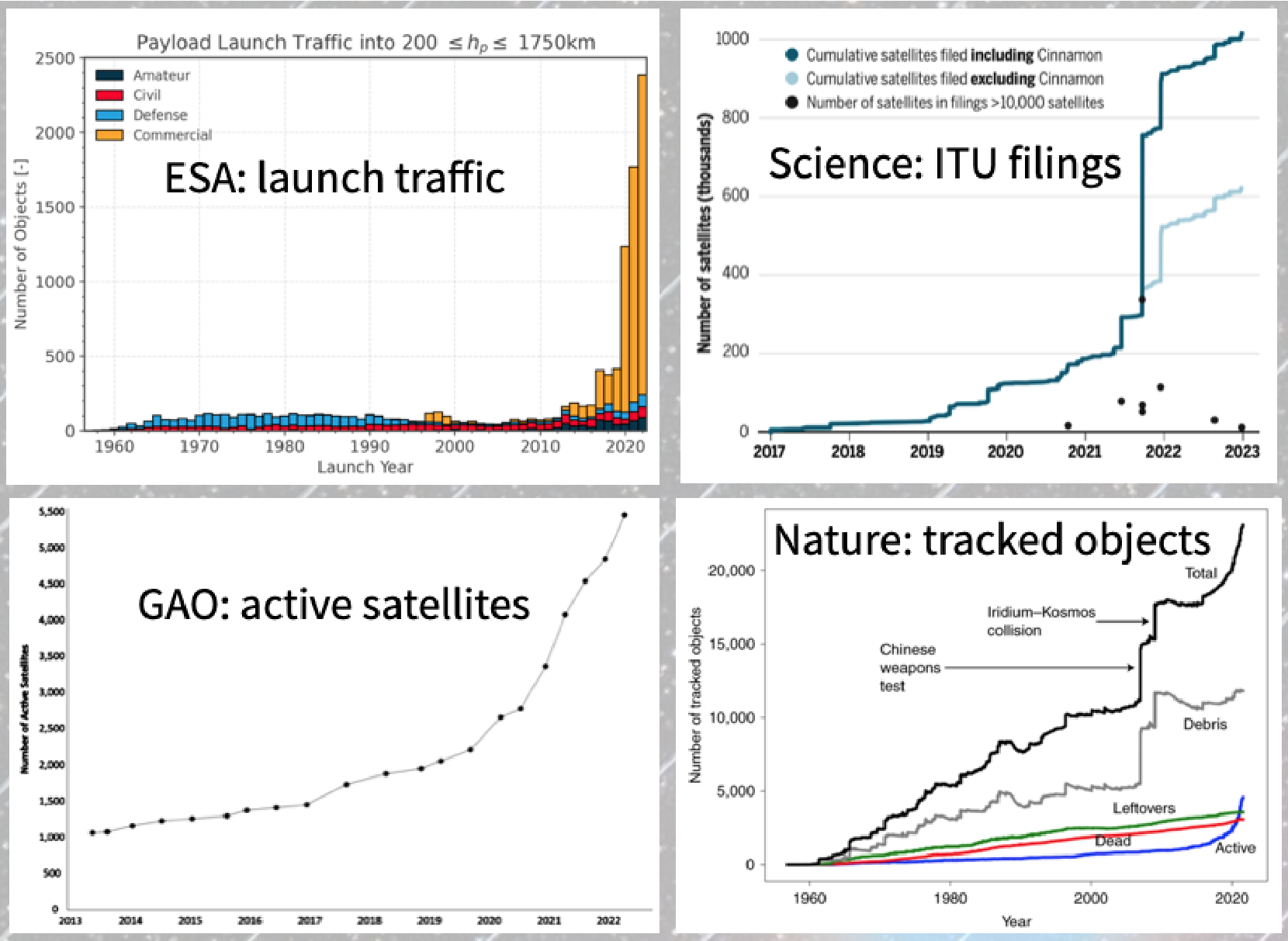}{fig1}{Four illustrations of the situation in low-Earth orbit. Each plot shows time in years on the x-axis. Upper left: launch traffic in number of objects to orbits between 200--1750 km; ``commercial'' is represented by yellow bars that overwhelmingly dominate the past three years \citep{esa23}. Upper right: the number of so-called ``paper satellites'', i.e., the number of satellites various companies and groups have filed an intention to launch with the ITU \citep{falle23}. Lower left: the number of active satellites, with the most recent data point around 5,500 \citep{gao22}. Lower right: the number of tracked objects in orbit, broken out into categories including debris, leftovers, dead and active satellites \citep{lawrence22}.}

In this proceedings, we summarize the discussions around two main topics: techniques and tools for forecasting satellite passes and approaches toward detecting and masking streaks in images. We close with a few other timely topics worthy of further discussion.

\section{Forecasting satellite positions and brightnesses for planning observations}
We introduced a new preliminary tool, SatChecker\footnote{\url{https://satchecker.readthedocs.io}}, that offers a publicly accessible API for predicting passes of a known satellite. Questions and discussions included how uncertainties are handled, what tools are available for forecasting glints or flares, which present-day observers are concerned about satellite constellations and what they are doing about it, and how radio astronomy can be incorporated.

We also highlighted that SatHub has completed and begun several successful observation campaigns toward quantifying the impacts of specific satellites or populations of satellites. The most recent example of this is AST SpaceMobile's BlueWalker 3 \citep{nandakumar23-bw3}. Other campaigns are planned for Starlink Gen2 Mini satellites --- several hundred have already been launched --- and the newly launched pair of Amazon Kuiper satellites, one with darkening mitigations and one without.

\section{Streak detection and masking algorithms and approaches}

In this segment, we first discussed how the Vera C. Rubin Observatory Legacy Survey of Space and Time (LSST) Science Pipelines approaches detecting and masking streaks \citep{saunders21}. The LSST Science Pipelines recently added the STREAK mask plane to all images run through standard single-frame processing pipelines, and validation using images that are known to contain satellite streaks is ongoing.

Conversation then turned to the potential of Zooniverse \citep[e.g., ][]{kruk23} or machine learning techniques to differentiate astrophysical sources from satellite artifacts, and what optical astronomers can learn from radio astronomers who are well-versed in radio frequency interference. This is a very new field of study and has many opportunities for novel contributions.

\section{Other concerns}
While the expertise of the authors is optical astronomy and software development, it is apparent that a good number of radio astronomers are highly engaged in this issue. For more on radio impacts, see, e.g., \citet{divruno23,grigg23,peel23-iaus385}.

In addition, IAU CPS Policy Hub Co-Lead (more recently appointed Interim Director) Richard Green described the slow, changing processes pertaining to regulations, policies, coordination agreements, etc., and ongoing work to get regulatory bodies to address aggregate effects rather than individual satellite impacts. For more on this and other broader impacts, see, e.g., \citet{green21-satcon2,venkatesan21-satcon2}.

We noted that significant work remains to understand how glints or flares from satellites or debris will affect different science investigations from surveys like LSST \citep{karpov23}.

Finally, we planned a followup meeting in 2023 December for those interested in forming a new SatHub Software Engineering Working Group. Several session attendees, including experts in planetary defense and radio astronomy, plan to be involved. The authors look forward to productive collaborations that identify existing techniques and unmet needs toward more effectively understanding the impacts of satellite constellations on astronomy.


\bibliography{B901.bib}

\end{document}